\documentclass[aps,prb,amssymb,twocolumn,showpacs]{revtex4}

\usepackage{graphicx}

\newcommand{\bea}{\begin{eqnarray}}

\newcommand{\eea}{\end{eqnarray}}

\newcommand{\beq}{\begin{equation}}

\newcommand{\eeq}{\end{equation}}

\newlength{\textwidthm}

\def \br{{\bf r}}

\def \bk{{\bf k}}

\def \bx{{\bf x}}
\def \bm{{\bf m}}

\begin{document}

\title{Phase fluctuations in anisotropic Bose condensates: from cigars to rings}

\author{L.~Mathey, A.~Ramanathan, K.~C.~Wright, S.~R.~Muniz,
 W.~D.~Phillips and Charles~W.~Clark}

\affiliation{Joint Quantum Institute, National Institute of Standards and Technology and University of Maryland,
 Gaithersburg, Maryland 20899, USA}

\date{\today}

\begin{abstract}
  We study the phase-fluctuating condensate regime of ultra-cold atoms
  trapped in a ring-shaped trap geometry, which has been realized in
  recent experiments.  We first
  consider a simplified box geometry, in which we identify the
  conditions to create a state that is dominated by thermal
  phase-fluctuations, and then explore the experimental ring 
  geometry.  In both cases we demonstrate that the requirement for
  strong phase fluctuations can be expressed in terms of the total number of
  atoms and the geometric length scales of the trap only.  For the
  ring-shaped trap we discuss the zero temperature limit in which a
  condensate is realized where the phase is fluctuating due to
  interactions and quantum fluctuations.  We also address possible
  ways of detecting the phase fluctuating regime in ring condensates.
\end{abstract}


\maketitle
Since the laboratory realization of Bose-Einstein condensates in
ultra-cold atomic systems, their properties in different dimensions
and geometries and the effect of quantum and thermal fluctuations have
been interesting questions.  In Refs. [\onlinecite{petrov}] and
[\onlinecite{oehberg}], it was pointed out that for an elongated,
`cigar'-shaped condensate, a regime of strong thermal fluctuations can
exist.  Such a system is three-dimensional in the sense that its
transverse dimensions are significantly larger than the healing length
and the thermal de Broglie wavelength.  However, the long-range
behavior of the single particle (field-field) correlation function
$g_1(\br) \equiv \langle\psi^\dagger(0)\psi(\br)\rangle$, with
$\psi(\br)$ being the single particle operator, can be dominated by
phase fluctuations along the longitudinal direction. This leads to an
exponential decay with some correlation length $l_\phi$, which is
typical for a one-dimensional system at finite (non-zero) temperature.
Because the many-body state combines these properties that are
characteristic for 3D and 1D systems, such a system can be considered
to be of mixed dimensionality.  When the correlation length $l_\phi$
is of the order of the system size or shorter, the condensate is said
to be in the phase-fluctuating regime.  In
Refs. [\onlinecite{khawaja}] and [\onlinecite{gerbier}] the 1D-3D
cross-over was studied.  In Ref. [\onlinecite{druten}], a two-step
condensation mechanism for the non-interacting system was pointed out.
The phase-fluctuating regime of elongated condensates was first
experimentally realized in Ref. [\onlinecite{dettmer}], and then
studied in further detail in Refs. [\onlinecite{hellweg, gerbier2003,
  richard, hugbart}].  We note that quantum fluctuations can lead to a
phase fluctuating condensate even at zero temperature, as we discuss
in Sect. \ref{zeroTemp}. However, here the decay of the correlation
function is algebraic, and we find that the exponent is very small for
the parameter regime of interest. The main focus of this article are
thermal phase fluctuations.

In this paper we study the properties of a ring-shaped
phase-fluctuating condensate.  One way to create such a system is by
using both a quasi-2D optical dipole trap formed from a sheet of light
(see Ref. [\onlinecite{clade}]), and a Laguerre-Gauss (LG) beam
perpendicular to the plane of the sheet [\onlinecite{details}].  In
such a setup, the atoms can form a condensate along the ring-shaped
maximum of the combined intensities of the sheet trap and the LG beam.
(A different approach was demonstrated in Ref. [\onlinecite{ryu}].)
Other realizations of toroidal traps were reported in
Refs. [\onlinecite{henderson}], [\onlinecite{perrin}] and
[\onlinecite{muniz}].  Here we study under what circumstances the
regime of thermal phase fluctuations can be reached for a condensate
with a ring geometry.

In Sect. \ref{box}, we first take the simplified case of a box
geometry with periodic boundary conditions, with different box sizes
 along different axes, with one of them much larger than the others.
In Sect. \ref{ring} we study the case of a ring-shaped condensate.  In
Sect. \ref{test} we consider ways in which this state can be seen
experimentally, and in Sect. \ref{conclude} we conclude.

\section{Box geometry}\label{box}
In this section we consider a condensate in a three dimensional box
 with periodic boundary conditions,
 and dimensions $X$, $Y$, and $Z$.  Our goal is to study the
behavior of the single-particle correlation function for the
dimensional cross-over from three dimensions to one dimension. This
correlation function can be used to identify the phase-fluctuating
regime of the condensate.

We consider the regimes $Z \approx X, Y$ and $Z \gg X, Y$, using the
calculational approach that is used in Refs. [\onlinecite{petrov}] and
 [\onlinecite{oehberg}].  This Bogoliubov-de Gennes (BdG) approach is
particularly useful for spatially inhomogeneous systems.  However, we
apply it to the box geometry as well, to allow for a clearer
 comparison between this case and the results
 for a ring geometry in Sect. \ref{ring}.
  We first calculate the BdG modes, and then consider the thermal occupation of these modes.

In a qualitative sense, the box geometry with
 periodic boundary conditions
 displays the main features of a Bose-Einstein condensate in elongated
 geometries. Because it is calculationally
 simpler, it will serve as the main system to understand
 the properties of fluctuating condensates.
 In Sect. \ref{ring} we derive the analogous properties
 for the spatially inhomogeneous case, and compare them to results
 of this section.

 We consider an atomic ensemble of bosons of density $n= N/V$, where
 $N$ is the total particle number, and the volume $V$ is $XYZ$.
 These atoms are assumed to be weakly interacting via a contact
 interaction with strength $g = 4\pi \hbar^2 a/M$, where $a$ is
 the scattering length, and $M$ is the atomic mass.  We then
 apply a BdG approach to describe the condensate and the excitations.
 The condensate is described by the well-known Gross-Pitaevskii (GP)
 equation, Ref. [\onlinecite{GPE}]:
\bea\label{GPEbox}
\Big(-\frac{\hbar^2\nabla^2}{2M}  +
 g |\Psi_0|^2\Big)\Psi_0 & = & \mu\Psi_0
\eea
where $\Psi_0$ is the condensate wave function, and $\mu$ is the
chemical potential.  Within the Thomas-Fermi (TF) approximation, see Ref. [\onlinecite{becreview}],
$\Psi_0$ is given by
\bea
\Psi_0 & = & \sqrt{n_{0}},
\eea
where $n_0=N_0/V$ is the condensate density, and $N_0$ is the
total number of atoms in the condensate.  
 For the weak interactions considered here, the quantum depletion is small and 
 $N_0 \approx N$.
The Bogoliubov-de Gennes
(BdG) equations [\onlinecite{degennesbook}] are given by
\bea
\Big(-\frac{\hbar^2\nabla^2}{2M}\Big)u_\nu +
 g |\Psi_0|^2(2 u_\nu - v_\nu) & = & (\mu + E_\nu)u_\nu\label{BdG1}\\
\Big(-\frac{\hbar^2\nabla^2}{2M}\Big)v_\nu +
 g |\Psi_0|^2(2 v_\nu - u_\nu) & = & (\mu - E_\nu)v_\nu.\label{BdG2}
\eea
The fields $u_\nu$ and $v_\nu$ describe an excitation of energy
$E_\nu$.  We note that for a phase-fluctuating condensate, this
formalism should be interpreted as an expansion in the density around
a finite value, and in the phase to quadratic order, see
Ref. [\onlinecite{mora}].  We introduce the rescaled quantities:
 $\bar{x} = x/X$, 
 $\bar{y} =  y/Y$,
 $\bar{z} = z/Z$,
 and
$\zeta^2  =  \hbar^2/(2 M \mu L^2)$
 where $L$ is defined as
$L  =  (X Y Z)^{1/3}$.
Our objective is to generate a consistent expansion in $\zeta$.
Written in these rescaled quantities, the GP equation (\ref{GPEbox}) is
given by
\bea\label{GPEboxrs}
-\zeta^2\tilde{\nabla}^2 \Psi_0
 + \bar{n}_0\Psi_0 & = & \Psi_0,
\eea
where $\tilde{\nabla}^2$ is defined as
\bea
\tilde{\nabla}^2 & = & \frac{L^2}{X^2}
 \frac{\partial^2}{\partial \bar{x}^2}
 + \frac{L^2}{Y^2}\frac{\partial^2}{\partial \bar{y}^2}
 +\frac{L^2}{Z^2} \frac{\partial^2}{\partial \bar{z}^2}.
\eea
The reduced density $\bar{n}_0$ is given by
  $\bar{n}_0  = g |\Psi_0|^2/\mu$.
 We note that the TF solution is simply the zeroth order solution in $\zeta$,
 i.e.  $\bar{n}_0 = 1$.
The rescaled BdG equations are 
\bea
-\zeta^2\tilde{\nabla}^2 u_\nu + \bar{n}_0 (2u_\nu - v_\nu)  & = &
 (1 + 2\zeta \epsilon_\nu)u_\nu\label{BdGred1}\\
-\zeta^2\tilde{\nabla}^2 v_\nu + \bar{n}_0 (2 v_\nu - u_\nu) & = &
 (1 - 2\zeta \epsilon_\nu) v_\nu,\label{BdGred2}
\eea
where  $2\zeta\epsilon_\nu = E_\nu/\mu$.
 We now define the fields $f_{\pm, \nu} \equiv u_\nu \pm v_\nu$.  These
fields are related to the phase and density fluctuations of excitation
$\nu$.  In terms of these fields, the  BdG equations (\ref{BdGred1}) and
 (\ref{BdGred2}) are
\bea
-\zeta^2\tilde{\nabla}^2 f_{+, \nu}
& = & 2\zeta \epsilon_\nu f_{-, \nu}\label{BdGf1}\\
-\zeta^2\tilde{\nabla}^2 f_{-, \nu}  + 2 f_{-, \nu}  & = &
 2\zeta \epsilon_\nu f_{+, \nu}.\label{BdGf2}
\eea
The solutions can be written as
$f_{\pm, \bm}  =  C_{\pm, \bm}  W(\bm)$.
%
 Here we replaced the generic index $\nu$ by the vector index $\bm$
that appears in the solutions.  The prefactors have to be related as
$C_{-, \bm} = \zeta \epsilon_\bm C_{+, \bm}$,
due to the normalization conditions on $u_\bm$ and $v_\bm$, see
Refs. [\onlinecite{petrov}] and [\onlinecite{oehberg}].
 The functions $W(\bm)$ are plane waves,
$W(\bm)  =  \exp(2\pi i \bm \bar{\br})$,
 where $\bm = (m_1, m_2, m_3)$, and $m_1, m_2$ and $m_3$ are integers.
 $\bar{\br}$ is defined as $\bar{\br} = (\bar{x},\bar{y},\bar{z})$.
  Eliminating $f_{-, \bm}$ from Eqs. (\ref{BdGf1}) and (\ref{BdGf2}), we can obtain the low-momentum limit of  energy-momentum dispersion 
   relationship 
\bea\label{eq11}
-\tilde{\nabla}^2 f_{+, \bm}
& = & 2 \epsilon_\bm^2 f_{+, \bm}.
\eea
%
%
%
 Written in terms of the original parameters, this gives
 $E_\bm   =  \hbar c |\bk|$,
 where the phonon velocity is
 $c  =  \sqrt{g n_0/M}$, and
$\bk$ is defined as $\bk = (2\pi m_1/X, 2\pi m_2/Y, 2\pi m_3/Z)$.
The normalization of the BdG modes is given by
 $C_{+, \bm}^2  = 2\mu/(E_\bm V)$.
In summary, the solutions of the BdG equations are
\bea
f_{+,\bm} & = & \sqrt{\frac{2\mu}{V E_\bm}}
 \exp(2\pi i \bm \bar{\br})\label{fplusbox}\\
f_{-, \bm} & = & \sqrt{\frac{E_\bm}{V 2\mu}}
 \exp(2\pi i \bm \bar{\br}).\label{fminusbox}
\eea
 Within the weak interaction expansion used here, we have $\mu = g n_0$.

 We now use these modes to calculate the correlation function of the
 phase.  The phase field $\phi(\br)$ is related to $f_{+, \bm}$
 through $\phi(\br) = 1/(2\sqrt{n_0})\sum_\bm f_{+,\bm} a_\bm^\dagger
 + h.c.$, where $a_\bm^\dagger$ is the bosonic creation operator of
 the $f_{+,\bm}$ mode.  The $f_{-, \bm}$ modes, on the other hand, are
 related to the density fluctuations of the system.  Because the modes
 $f_{+, \bm}$ behave as $1/\sqrt{E_\bm}$, whereas $f_{-, \bm}$ scale
 as $\sqrt{E_\bm}$, one can expect that at long distances (small $k$) 
  the phase of the system can be
 a fluctuating quantity, whereas the density fluctuations will be
 suppressed.

 The correlation function of the phase is given by
\bea\label{phicorr}
\langle \delta\phi(\br)^2\rangle & = & \sum_\bm \frac{g}{V E_\bm} N_\bm
 (2 -2 \cos(2\pi\bm \bar{\br})),
\eea
where $\delta\phi(\br) \equiv \phi(0) - \phi(\br)$.
We assume that the Bogoliubov modes are occupied according to a
 Bose-Einstein distribution of temperature $T$.  Thus, the distribution $N_\bm$,
which is the occupation of a state with energy $E_\bm$, is given by
\bea\label{Nm}
N_\bm & = & \frac{1}{\exp(E_\bm/k_B T) -1} + 1/2,
\eea
 where the $1/2$ accounts for the vacuum fluctuations. 
For small energies and at finite temperature, $N_\bm$ has a divergent
behavior, which is the origin of the phase-fluctuating regime. It
behaves as
\bea\label{Napprox}
N_\bm & \approx & \frac{k_B T}{E_\bm}.
\eea
This divergent term dominates over the constant term $1/2$, which we
discuss in Sect. \ref{zeroTemp}. We find that it generates an
algebraically decaying correlation function, with an exponent that is
very small for the parameter regime of interest.  In particular, the
term does not contribute to the exponential decay, and will be ignored
in this section.

\subsection{Correlation function in 1D and 3D}
The correlation function for the elongated
 geometry shows 3D behavior on short length scales $r\lesssim X, Y$,
 and 1D behavior for longer scales $r \gtrsim X, Y$.
 In this section we discuss these limits.

We first calculate the 3D limit by replacing the sums in
Eq.~(\ref{phicorr}) with integrals.  The thermal contribution of the
correlation function is
\bea\label{phicorr3D}
\langle \delta\phi(\br)^2\rangle & = & \frac{g}{2\pi^2 \hbar c}
\Big(\frac{\pi^2(k_B T)^2}{3 \hbar^2 c^2}\nonumber\\
&& + \frac{1}{r^2}-\frac{\pi k_B T\coth(\pi r k_B T/\hbar c)}{\hbar c r}\Big),
\eea
where $r=|\br|$.  For distances $r$ large compared to the healing
length $\xi =\hbar/(2 M g n_0)^{1/2}$ and the thermal de Broglie
wavelength $\lambda = \hbar/(2\pi M k_B T)^{1/2}$, this approaches
\bea\label{phicorr3Dapprox}
\langle \delta\phi(\br)^2\rangle & \approx & \frac{g}{2\pi^2 \hbar c}
\Big(\frac{\pi^2(k_B T)^2}{3 \hbar^2 c^2}
-\frac{\pi k_B T}{\hbar c r}\Big),
\eea
(see e.g. Ref. [\onlinecite{rice}]).  This expression shows the $1/r$
decay of the phase correlations of a 3D BEC to a non-zero value,  related to the thermal
depletion at $T \lesssim \mu$. The thermal depletion is  proportional to $\xi/(n\lambda^4)$.
Equation (\ref{phicorr3Dapprox})
 is valid in the
 elongated geometry for $\xi, \lambda \ll r \lesssim X, Y$.

As we approach the 1D limit, for  $r \gg X, Y$, the modes with $m_1=m_2=0$ produce the dominant
contribution to the correlation function, whereas the
 excited  transverse modes provide corrections, which we
discuss below.  We write the correlation function
 as  $\langle\delta\phi(\br)^2\rangle =
\langle\delta\phi(\br)^2\rangle_0 + \langle\delta\phi(\br)^2\rangle_1$, where
$\langle\delta\phi(\br)^2\rangle_0$ corresponds to the modes with $m_1 =
m_2 = 0$, and $\langle\delta\phi(\br)^2\rangle_1$ to the remaining
contributions in Eq. (\ref{phicorr}).  For
$\langle\delta\phi(\br)^2\rangle_0$, with the approximation in
Eq. (\ref{Napprox}), we find
\bea
\langle \delta\phi(\br)^2\rangle_0 & = &  \frac{g k_B T}{X Y}
 \frac{Z}{(\hbar c)^2} (|\bar{z}|-\bar{z}^2).
\eea
 In this expression, the linear term contains the length scale that we are
 interested in, whereas the quadratic term ensures that the
 derivative of the function at $\bar{z} = 1/2$ is smooth.
 One might expect
 an expression that is purely given by trigonometric functions of $\bar{z}$,
 however,
  the approximation in Eq. (\ref{Napprox}) leads to the expression above.
 We note that if we take the limit $Z \rightarrow \infty$,
 the linear term stays fixed, because it is only a function of $z$,
 whereas the quadratic term vanishes, because it scales as $1/Z$.

\begin{figure}
\includegraphics[width=6.2cm]{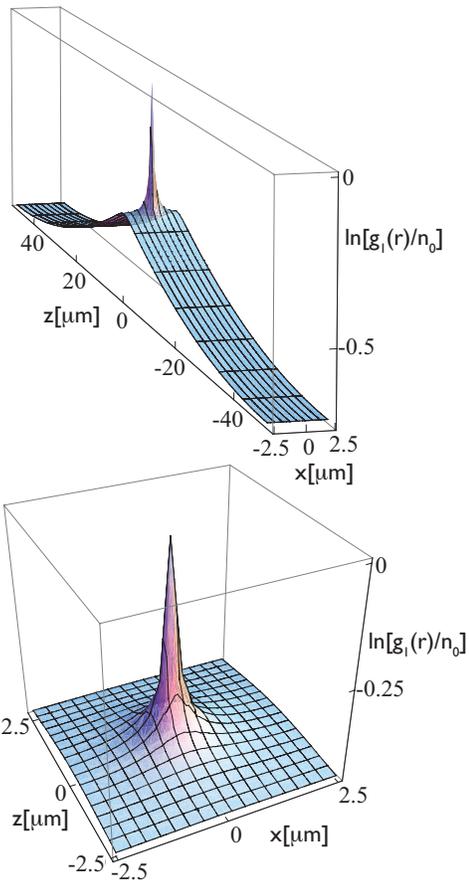}
\caption{\label{corr} (Color online) Thermal contribution to the
  correlation function $g_1(\br) =
  \langle\psi^\dagger(0)\psi(\br)\rangle$.  We plot $\ln(\langle
  \psi^\dagger(0)\psi(\br)\rangle/n_0)$, which equals $-\langle
  \delta\phi(\br)^2 \rangle/2$.  We use a box geometry with $Z=100
  \mu$m and $X=Y=5 \mu$m, and a temperature $T=100$nK.  The total atom
  number is $N_0 = 20000$, and $M$ is the sodium mass $M=23 u$, with
  $u$ being the atomic mass unit.  We assume that the scattering
  length is $a = 2.75$nm.  The upper diagram shows the full
  correlation function, the lower diagram the regime $z \lesssim X,
  Y$.  }
\end{figure}

 At intermediate distances $X,Y \ll r \ll Z$ (where the periodic
 boundary conditions are not apparent), the single particle
 correlation function thus approaches
\bea
\langle \psi^\dagger(0) \psi(\br)\rangle & \sim & n_0
 \exp(- Z |\bar{z}|/l_\phi),
\eea
 where we used the phase-density representation
 of the boson operator $\psi(\br) = \sqrt{n(\br)}\exp(i\phi(\br))$
 to connect the single particle correlation function to the
 phase correlation function.
  The phase correlation length $l_\phi$ is given by
\bea\label{lphibox}
l_\phi & = & \frac{2 X Y (\hbar c)^2}{g k_B T}
 =  \frac{2 X Y \hbar^2 n_{0}}{M k_B T}
 =  \frac{2 \hbar^2 N_0}{M Z k_B T}.
\eea
 We note that the interaction has cancelled in this expression
 (see e.g. Refs. [\onlinecite{gerbier}] and [\onlinecite{cazalilla}]).
  This can be understood by writing the Hamiltonian in the
 phase-density representation. As discussed in Ref. [\onlinecite{mora}],
 the Hamiltonian of a weakly interacting Bose gas is given to second
 order by a kinetic term $\sim n_0 (\nabla \phi)^2/2M$ and a potential
 term of the form $g \delta n^2$, where $\delta n$ are the density
 fluctuations around the constant value $n_0$.  In the momentum
 representation, the spectrum of this Hamiltonian can be seen to be
 equivalent to the Bogoliubov spectrum of Eqs. (\ref{BdG1}) and
 (\ref{BdG2}).
 The contribution
 to the
 partition function made by
  a given mode of energy $\hbar \omega_k$
 consists of independent phase
 and density terms, for a temperature $T \gtrsim \hbar \omega_k/k_B$.  In this
 limit, the phase correlations are entirely due to the kinetic term,
 leading to the expression in Eq. (\ref{lphibox}), independent
 of the interaction strength.

 Next, we estimate the effect of the higher modes, for the case
 $X = Y$. We consider the thermal contribution to the spatially
 independent part of the correlation function.  We replace the $m_3$
 summation in Eq. (\ref{phicorr}) by an integral over $k_z$, and expand the dispersion relation appearing in the 
  exponent in Eq. (\ref{Nm}) to
 second order in $k_z$.  We can then evaluate the $k_z$
 integral as
\bea
\langle \delta\phi(\br)^2\rangle_1 & \approx &
 \sum_{m_1, m_2 \neq 0} \int \frac{d k_z}{2\pi} \frac{g}{\pi \hbar c X
  (m_1^2 + m_2^2)^{1/2}}\nonumber\\
 && \exp\Big(-\frac{8\pi^2 \hbar c (m_1^2 + m_2^2) +\hbar c X^2 k_z^2}{4 \pi X
  k_B T (m_1^2 + m_2^2)^{1/2}}\Big)\nonumber\\
& = &
 \sum_{m_1, m_2 \neq 0} \frac{g}{\pi \hbar c X (m_1^2 + m_2^2)^{1/4}}
\Big(\frac{k_B T}{X \hbar c}\Big)^{1/2}\nonumber\\
 && \exp\Big(-\frac{2\pi \hbar c (m_1^2 + m_2^2)^{1/2}}{X k_B T}\Big).
\eea
Next we approximate the sum $\sum_{m_1, m_2 \neq 0}$ by
$\sum_{m=1}^{\infty} 2\pi m$, and $(m_1^2 + m_2^2)^{1/2}$ by $m$. That
gives
\bea
\langle \delta\phi(\br)^2\rangle_1 & \approx &
 \sum_{m} \frac{2 g \sqrt{m}}{\hbar c X} \Big(\frac{k_B T}{X \hbar c}\Big)^{1/2}
 \exp\Big(-\frac{2\pi \hbar c m}{X k_B T}\Big)\nonumber\\
 &=&  \frac{2 g (k_B T)^{1/2}}{(\hbar c X)^{3/2}}
 Li_{-\frac{1}{2}}\Big[\exp\Big(-\frac{2\pi \hbar c}{X k_B T}\Big)\Big]\label{phi23}
\eea
where $Li_{-\frac{1}{2}}$ is the standard polylogarithmic function with
exponent $-1/2$.

For temperatures large enough to populate transverse modes, i.e.
 $k_B T\gg 2\pi \hbar c/X$, but smaller than or
 comparable to $\mu$,
 Eq. (\ref{phi23}) becomes 
\bea\label{depletion}
\langle \delta\phi(\br)^2\rangle_1 & \approx &
  \frac{g (k_B T)^2}{2^{3/2}\pi(\hbar c)^3}.
\eea
We therefore approximately recover the 3D thermal depletion, which we
found before in Eq. (\ref{phicorr3Dapprox}).  For $k_B T \ll 2\pi \hbar c/X$, 
 $\langle \delta\phi(\br)^2\rangle_1$ is exponentially suppressed
\bea
\langle \delta\phi(\br)^2\rangle_1 & \approx &
  \frac{2 g}{\hbar c X} \Big(\frac{k_B T}{X \hbar c}\Big)^{1/2}
 \exp\Big(-\frac{2\pi \hbar c}{X k_B T}\Big).
\eea
In Fig. \ref{corr} we show the full thermal contribution to the
correlation function, Eq. (\ref{phicorr}), for box dimensions $Z = 100
\mu$m and $X = Y = 5 \mu$m.  The temperature is $T = 100$nK, and the
atomic mass is $M = 23 u$, with $u$ being the atomic mass unit.  For $z
\lesssim X, Y$ the correlation function $g_1(\br)$ closely resembles the 3D
result (see Eq. (\ref{phicorr3D})), i.e.  it falls off on a short scale to a
finite value, which describes the thermal depletion of the condensate.
For sufficiently large $z \gg X, Y$, the correlation function $g_1(\br)$ is approximately
 given by $g_1(z)$, and  behaves like the correlation function for a 1D system.
The thermally activated modes where $m_1, m_2 = 0$, $m_3\neq 0$ generate an
exponentially decaying $g_1(z)$.  In this 1D-like case, the
correlation function is essentially only a function of the coordinate
$z$, and independent of $x$ and $y$.  The effect of the transverse
modes is to create a constant thermal depletion, as in
Eq. (\ref{depletion}), which is of the order of the 3D result.  So the
cross-over from 3D to 1D is given by the geometric shape, and the
regimes are $z\lesssim X, Y$ for 3D behavior, and $z \gtrsim X, Y$ for
1D behavior (see also Ref. [\onlinecite{mikeska}]).

\begin{figure}
\includegraphics[width=8cm]{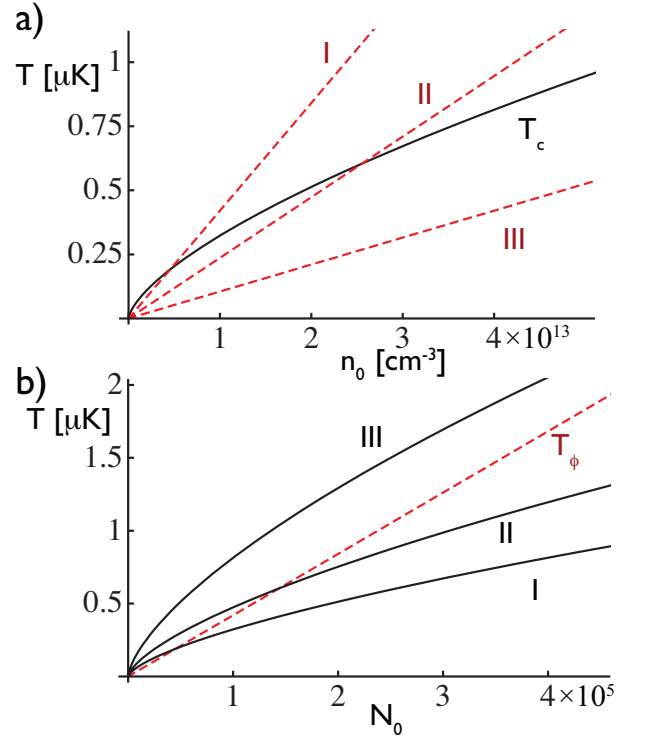}
\caption{\label{TphiTc} (Color online) Comparison of $T_c$ (black,
  continuous lines) and $T_\phi$ (red, dashed lines), for different
  geometric aspect ratios. In a) we show these temperatures as a
  function of density, for the sodium mass $M=23 u$, and $Z=100 \mu$m,
  and for $X=Y=10, 7.5, 5 \mu$m (I -- III).  In b) we show the same
  quantities as a function of the total atom number $N_0$. $T_c$ is
  purely a function of the density, whereas $T_\phi$ is a function of
  the total number, for a fixed longitudinal dimension $Z$.}
\end{figure}
%

\subsection{Phase-fluctuating condensate}
In the previous section we found that for an elongated system the long
range behavior of the single-particle correlation function $g_1(\br)$ shows
exponential decay with a correlation length $l_\phi$.  We can then
define a temperature above which the system is in the
phase-fluctuating regime, based on the requirement that this length
scale is comparable to the system size, i.e. $l_\phi \approx Z$.
This defines the following temperature
\bea
k_B T_\phi & = & \frac{2 \hbar^2 N_0}{M Z^2}.
\eea
  If this temperature
 lies below the BEC temperature $T_c$,
 a phase-fluctuating condensate may exist.
 We estimate that temperature by using the
 result for the homogeneous, ideal Bose gas, which is (e.g. Ref. [\onlinecite{pethicksmith}])
\bea
k_B T_c & = & \frac{2\pi}{\zeta(3/2)^{2/3}} \frac{\hbar^2 n_0^{2/3}}{M}, 
\eea
 where $\zeta$ signifies the Riemann zeta function.   
To estimate the regime in which one would observe a phase-fluctuating
 condensate, we consider the ratio of $T_c$ and $T_\phi$,
\bea
T_c/T_\phi & = & \frac{\pi}{\zeta(3/2)^{2/3}}
 \frac{1}{N_0^{1/3}} \frac{Z^{4/3}}{X^{4/3}},
\eea
 which we wrote in terms of the total
 atom number $N_0$, and we chose $X=Y$.
  For a system of fixed dimensions $X$ and $Z$, 
   no phase-fluctuating regime exists if
   the total
 number of atoms  fulfills
\bea\label{N0}
N_0 &\gtrsim& \frac{\pi^3}{\zeta(3/2)^{2}} \frac{Z^4}{X^4} \approx
 4.5 \frac{Z^4}{X^4},
\eea
demonstrating how sensitively the presence or absence depends on the geometric
ratio $Z/X$.  In Fig. \ref{TphiTc} we show both the condensate
temperature and $T_\phi$ as a function of density, for different
geometric dimensions. The regime above $T_\phi$ but below $T_c$ is the
phase-fluctuating condensate regime.  We can again see that the
magnitude of this regime is very sensitive to the ratio $Z/X$.

\section{Ring geometry}\label{ring}
We now turn to the kind of ring-shaped geometry in which we are
interested.  As mentioned in Sect.~\ref{box}, we will find that
qualitatively most features of the phase-fluctuating condensate in a
ring can be understood in terms of the simplified box geometry
discussed previously.  We note that the elongated box with periodic
boundary condition is very similar to a ring.
 In addition, the cigar-shaped condensates discussed in 
 Refs.~[\onlinecite{petrov}] and [\onlinecite{oehberg}] qualitatively
 share similar features in the center of the condensate.
  In this section, we
derive the results analogous to those of the previous section but for
a trapped (inhomogeneous) ring system and compare them to results of
Sect. \ref{box}.

\begin{figure}
\includegraphics[width=7cm]{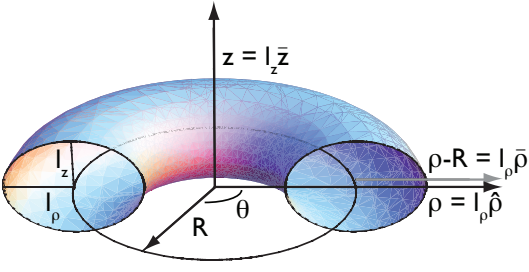}
\caption{\label{coor} (Color online) The toroidal shape of the BEC is
  given by the radius $R$, and the Thomas-Fermi radii $l_z$ in $z$
  direction, and $l_\rho$ in radial direction.  We use the rescaled
  $z$ variable $\bar{z} = z/l_z$, and the two rescaled radial
  variables $\hat{\rho} = \rho/l_\rho$ and $\bar{\rho} = (\rho -
  R)/l_\rho$, as indicated in the figure.  }
\end{figure}

 In the treatment here, the trapping potential is approximated
 by the harmonic form
\bea
V & = & \frac{1}{2} M \omega_z^2 z^2 + \frac{1}{2} M \omega_\rho^2
(\rho-R)^2.
\eea
 Here, $z$ and $\rho$ are spatial cylindrical
coordinates as shown in Fig. \ref{coor}, and $\omega_z$ and $\omega_\rho$ the oscillation frequencies along the spatial
 directions, and $R$ is the mean ring radius.  We apply the
same formalism as in Sect. \ref{box}.  The GP equation of the system is given
by
\bea\label{GPEring}
\Big(-\frac{\hbar^2\nabla^2}{2M} + V +
 g |\Psi_0|^2\Big)\Psi_0 & = & \mu\Psi_0,
\eea
 the BdG equations are
\bea
\Big(-\frac{\hbar^2\nabla^2}{2M} + V\Big)u_\nu & + &
 g |\Psi_0|^2(2 u_\nu - v_\nu)\nonumber\\
 && =  (\mu + E_\nu)u_\nu\\
\Big(-\frac{\hbar^2\nabla^2}{2M} + V\Big)v_\nu & + &
 g |\Psi_0|^2(2 v_\nu - u_\nu)\nonumber\\
 && =  (\mu - E_\nu)v_\nu.
\eea
The Thomas-Fermi (TF) solution of Eq. (\ref{GPEring}) for $\bar{z}^2+\bar{\rho}^2<1$ is
\bea
\Psi_0 & = & \sqrt{n_{\mathrm{max}} (1- \bar{z}^2 - \bar{\rho}^2)}.
\eea
Here we have rescaled the spatial coordinates $\bar{z} = z/l_z$ and
$\bar{\rho} = (\rho-R)/l_\rho$ by the TF radii, given by
 $l_z = (2\mu/M\omega_z^2)^{1/2}$ and $l_\rho =
 (2\mu/M\omega_\rho^2)^{1/2}$.  This rescaling is in analogy to
 rescaling the spatial coordinates of the box geometry by the
 dimensions of the box.  $n_{\mathrm{max}}$ is the maximum value of
 the density of the condensate.  We write the GP equation as
\bea
-\eta^2\tilde{\nabla}^2 \Psi_0 + \bar{z}^2 \Psi_0
+ \bar{\rho}^2\Psi_0 + \bar{n}_0\Psi_0 & = & \Psi_0
\eea
where we introduced the operator
\bea
\tilde{\nabla}^2 & = & \Big(\frac{\omega_\rho^2}{\bar{\omega}^2}
 \Big(\frac{\partial^2}{\partial\hat{\rho}^2}
 + \frac{1}{\hat{\rho}}\frac{\partial}{\partial \hat{\rho}} +
\frac{1}{\hat{\rho}^2}\frac{\partial^2}{\partial\theta^2}\Big)
 +\frac{\omega_z^2}{\bar{\omega}^2} \frac{\partial^2}{\partial \bar{z}^2}\Big)\label{nablasqrd},
\eea
 where 
  $\theta$ is the azimuthal angle, as shown in Fig. \ref{coor},   
 and $\hat{\rho} = \rho/l_\rho$.
 The frequency $\bar{\omega} =
\sqrt{\omega_z\omega_\rho}$,  and $\eta^2 =
\frac{\hbar\bar{\omega}}{2\mu}$ [\onlinecite{etazeta}].

 We now introduce a rescaled energy $\epsilon_\nu
= E_\nu/\hbar\bar{\omega}$ and  a rescaled density $\bar{n}_0$, 
defined as $\bar{n}_0 = |\Psi_0|^2/n_{\mathrm{max}}$.  
 With these parameters, 
 the BdG equations are
\bea
-\eta^2\tilde{\nabla}^2 u_\nu + (\bar{z}^2
+ \bar{\rho}^2) u_\nu & + & (2u_\nu - v_\nu) \bar{n}_0\nonumber\\ 
& & =  (1 + 2\eta \epsilon_\nu)u_\nu\\
-\eta^2\tilde{\nabla}^2 v_\nu + (\bar{z}^2
+ \bar{\rho}^2) v_\nu & + & (2 v_\nu - u_\nu) \bar{n}_0\nonumber\\
 & & =  (1 - 2\eta \epsilon_\nu) v_\nu.
\eea
These equations are analogous to Eqs. (\ref{BdGred1}) and (\ref{BdGred2}),
but here the equations contain a spatially dependent potential and
condensate density $\bar{n}_0 = \bar{n}_0(\bar{z}, \bar{\rho})$.

As before, we introduce the fields $f_{\pm, \nu} = u_\nu \pm v_\nu$.  In terms of
these, the BdG equations are
\bea
\Big(-\eta^2\tilde{\nabla}^2
& + &\frac{\eta^2\tilde{\nabla}^2\Psi_0}{\Psi_0}\Big)f_{+, \nu}\nonumber\\
& = & 2\eta \epsilon_\nu f_{-, \nu}\\
\Big(-\eta^2\tilde{\nabla}^2 & + & 2 - 2 \bar{z}^2
- 2 \bar{\rho}^2  +  
 \frac{3\eta^2\tilde{\nabla}^2\Psi_0}{\Psi_0}\Big)f_{-, \nu}\nonumber\\ 
& = &
 2\eta \epsilon_\nu f_{+, \nu}
\eea
 We write the solutions in the TF approximation as
\bea
f_{+, \nu} & = & C_{+, \nu} (1- \bar{z}^2 - \bar{\rho}^2)^{1/2} W_\nu(\theta)\\
f_{-, \nu} & = & C_{-, \nu} (1- \bar{z}^2 - \bar{\rho}^2)^{-1/2}W_\nu(\theta)
\eea
with the normalization condition $C_{-, \nu} = \eta \epsilon_\nu C_{+,
  \nu}$.
The functions $W_\nu(\theta)$ are simply $W_\nu = \exp(i m \theta)$,
with $m$ being an integer.  We replace the index $\nu$ with $m$
from now on.  For these solutions we assumed that the radius $R$ is
much larger than $l_\rho$, so that we can approximate the term
$(1/\hat{\rho}^2)\partial_\theta^2$ in Eq. (\ref{nablasqrd}) by
$(l_\rho^2/R^2)\partial_\theta^2$.

This set of modes is analogous to the modes with $m_1 = m_2 = 0$ in
the previous section.
 In analogy to Eq. (\ref{eq11}), the dispersion relation can be calculated from
\bea
(1 - \bar{z}^2 - \bar{\rho}^2)
\Big(-\tilde{\nabla}^2 +\frac{\tilde{\nabla}^2\Psi_0}{\Psi_0}\Big)f_{+, m}
 & = & 2 \epsilon_m^2 f_{+, m}
\eea
where we need to average over the density profile, see
Refs. [\onlinecite{petrov}] and [\onlinecite{stringari}].  This gives
 $E_m  =  \hbar c |k|$,
with $c = \sqrt{g n_{\mathrm{max}}/2M}$ and $k = m/R$.
 These modes and dispersion relation were also discussed within
 a hydrodynamic approach in Ref. [\onlinecite{stringari}].
 The prefactor $C_{+, m}$ is given by
 $C_{+, m}^2  =  1/(2\pi^2\epsilon_m \eta R l_z l_\rho)$,
 so $f_{\pm, m}$ are
\bea
f_{+,m}
 & = & \sqrt{\frac{g n_0(z,\rho)}{\pi^2 R l_z l_\rho E_m}}
 \exp(i m \theta)\\
f_{-, m} & = & \sqrt{\frac{E_m}{4 \pi^2 R l_z l_\rho g n_0(z,\rho)}}
 \exp(i m \theta).
\eea
These expressions are analogous to Eqs. (\ref{fplusbox}) and
(\ref{fminusbox}).  The `Thomas-Fermi' 
 volume $V$ in Eqs. (\ref{fplusbox}) and (\ref{fminusbox})
corresponds to $2\pi^2 R l_z l_\rho$, which is the volume of the
torus.  The chemical potential $\mu$ is replaced by $g n_0(\rho,z)$.

 Next, we calculate the correlation function of the phase
\bea
\langle \delta\phi(\theta)^2\rangle & = & \sum_m \frac{g N_m}{2\pi^2 R l_z l_\rho E_m}
 (2 -2 \cos(m\theta))
\eea
where $N_m$ is given in Eq. (\ref{Nm}). In analogy with $\delta\phi(\br)$ in the previous
 section,  we have defined 
 $\delta\phi(\theta) = \phi(0) - \phi(\theta)$.
 Using Eq. (\ref{Napprox}), we find
\bea
\langle \delta\phi(\theta)^2\rangle & = &  \frac{g k_B T}{\pi^2 l_z l_\rho}
 \frac{R}{(\hbar c)^2} (-\theta^2/2 + \pi |\theta|),
\eea
 in analogy to Eq. 19.
 So for intermediate distances $l_\rho, l_z \ll R\theta \ll R$  we have
 $\langle \psi^\dagger(\theta) \psi(\theta')\rangle  \sim
 \exp(- R |\Delta\theta|/l_\phi)$,
 with
\bea
l_\phi & = & \frac{\pi l_z l_\rho \hbar^2 n_{\mathrm{max}}}{M k_B T}
  =  \frac{\hbar^2 N_0}{\pi M R k_B T},
\eea
in analogy to Eq. 20.
 Here we used the relationship between the density maximum $n_{\mathrm{max}}$
and the total atom number $N_0$, $N_0 = n_{\mathrm{max}}\pi^2 l_z l_\rho R$.
Again we find that the interaction does not affect $l_\phi$.  We note
that this expression is of the same form as the expression for the box
geometry, Eq. (\ref{lphibox}), up to numerical prefactors.

\subsection{Phase-fluctuating regime}
 As in Sect. \ref{box}, the phase-fluctuating regime
 is reached when the phase correlation length
 is of the order of the system size.
 From $l_\phi \approx \pi R$ we can define the temperature
\bea\label{Tphiring}
k_B T_\phi & = &   \frac{\hbar^2 N_0}{\pi^2 M R^2}.
\eea
 We calculate  the condensation temperature for a cylinder
 of length $2 \pi R$ and with harmonic confinement with frequencies
 $\omega_z$ and $\omega_\rho$
\bea\label{Tcring}
k_B T_c & = & \Big(\frac{3 N_0 \hbar^3 \omega_z \omega_\rho}{4 \Gamma(\alpha)\zeta(\alpha) (2 m)^{1/2}R}\Big)^{2/5}
\eea
 where $\zeta(\alpha)$ 
 is again the Riemann function at $\alpha=5/2$,
  and $\Gamma(\alpha)$ is the usual Gamma function.

 As discussed in Ref. [\onlinecite{bagnato}], at this critical temperature 
  the density at the potential minimum
 reaches the critical value of the homogeneous system,
 within a semi-classical approximation. 

\begin{figure}
\includegraphics[width=8cm]{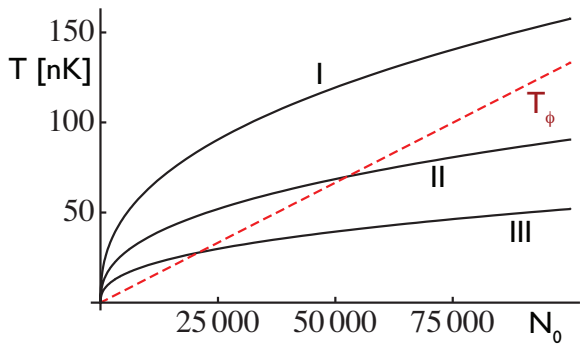}
\caption{\label{TcTphi} (Color online) $T_\phi$ (red, dashed line) as
  a function of the total atom number, for $M=23 u$ and $R=40 \mu$m,
  and $T_c$ (black, continuous lines) for
  $\omega_z=\omega_\rho=2\pi\times 200, 2\pi\times 100, 2\pi\times
  50$Hz (I -- III).  }
\end{figure}
 The ratio $T_c/T_\phi$ can be written as
\bea
\frac{T_c}{T_\phi} & = & \Big(\frac{3\pi^5}{4\sqrt{2}\Gamma(\alpha)\zeta(\alpha)}\Big)^{2/5}
 \frac{1}{N_0^{3/5}}\Big(\frac{R^2}{a_z a_\rho}\Big)^{4/5}
\eea
where $a_z =\sqrt{\hbar/m\omega_z}$ and $a_\rho =\sqrt{\hbar/m\omega_\rho}$
 are the oscillator lengths of the confining potential.
 For the total atom number this leads to the
 inequality
\bea
N_0 & \gtrsim & \Big(\frac{3\pi^5}{4\sqrt{2}\Gamma(\alpha)\zeta(\alpha)}\Big)^{2/3}
 \Big(\frac{R^2}{a_z a_\rho}\Big)^{4/3}\\
& \approx & 20.2 
 \Big(\frac{R^2}{a_z a_\rho}\Big)^{4/3},
\eea
for which there is no phase-fluctuating condensate for any temperature.
 This conclusion is valid if the
 estimate of the condensate temperature in Eq. (\ref{Tcring}) is accurate,
 and under the weak interaction conditions of the BdG approach.

In Fig. \ref{TcTphi} we show
$T_\phi$ and $T_c$ as a function of the total atom number $N_0$, and
for different confining frequencies $\omega_z$ and $\omega_\rho$.  For
a given $N_0$ and trap radius $R$, $T_\phi$ is fixed.  The
condensation temperature, however, is affected by the transverse 
confinement through $\omega_z$ and $\omega_\rho$.  By increasing their
value, a large regime of a phase-fluctuating condensate can be
created.
 On the other hand, for sufficiently weak transverse confinement,
 one can avoid the phase fluctuating regime for all temperatures.

\subsection{Zero temperature}\label{zeroTemp}
While one might expect that in 3D a condensate would have perfect
phase coherence at zero temperature, we find that the
phase-fluctuating character of an elongated system persists even at
zero temperature, as for a truly 1D system
[\onlinecite{giamarchi_book}]. At zero temperature, the single-particle
correlation function of our ring shaped condensate decays algebraically:
\bea
 \langle \psi^\dagger(\br_1)\psi(\br_2) \rangle & \sim &
\Big(\frac{(R \theta)^2}{r_c^2}\Big)^{-1/4K}\label{corrzero},
\eea
where $\br_1$ and $\br_2$ both are at the minimum of the trapping
potential, with $|\br_1| = |\br_2| = R$, and $\theta$ is the angle
between them.  Equation (\ref{corrzero}) is valid for $l_z, l_\rho,
r_c \ll |R \theta| \ll R$, where $r_c$ is a short-range cut-off which
should be chosen to be of the order of the healing length $l_h =
\frac{\hbar}{\sqrt{m \mu}}$, see [\onlinecite{cutoff}].

 The scaling exponent $K$ is given by
\bea
K & = & \frac{\hbar N_0^{3/4}}{2^{1/4}R^{3/4} M^{3/4}}
\frac{\pi^{1/2}}{g^{1/4} (\omega_z\omega_\rho)^{1/4}}.
\eea
For typical parameters, $K$ is very large.
 For example, for $R=40\mu$m, $M=23 u$, $a=2.75nm$, $\omega_z = \omega_\rho =
 2\pi \times 1000$Hz, and $N_0 = 20000$, we have $K =242.4$.
  Since the algebraically decaying correlation function reaches, say, the
 value $1/e$ at a distance $r_c e^{2K}$, which
 vastly exceeds the size of the system,  
  the decay of the
 correlation function across the system is very small.
 So although the system in principle forms a quasi-condensate, for
 typical parameters it will behave essentially like a true condensate.

%
%

\section{Experimental Signatures}\label{test}
In this section we consider possible experimental signatures of the
phase-fluctuating regime.  The defining feature of this regime is the
loss of phase coherence over the ring-shaped condensate.
 One way of testing this would be to use the
 interference technique demonstrated in Ref.~[\onlinecite{clade}].

In Ref.~[\onlinecite{dettmer}] the signature of the
 phase-fluctuating regime in time-of-flight images was studied.
 In Ref.~[\onlinecite{hellweg}] an interferometric approach measuring
 the second-order correlation function was reported. 
 In Ref.~[\onlinecite{richard}] Bragg spectroscopy was
 used to measure the momentum distribution in the axial
 direction, which was calculated in Ref.~[\onlinecite{gerbier2003}].
 Matter wave interferometry was used in Ref.~[\onlinecite{hugbart}]
 to measure the phase coherence length.

 A consequence of the ring geometry is that during a time-of-flight
 (TOF) expansion of the phase-fluctuating condensate,   
  the contrast of    
  matter-wave interference should be reduced compared
 to a true condensate.
 In particular, 
 due to the ring-shaped geometry of the system one can expect that the
 constructive interference that can emerge at the center
 of the ring during expansion will be reduced.

\begin{figure}
\includegraphics[width=8.7cm]{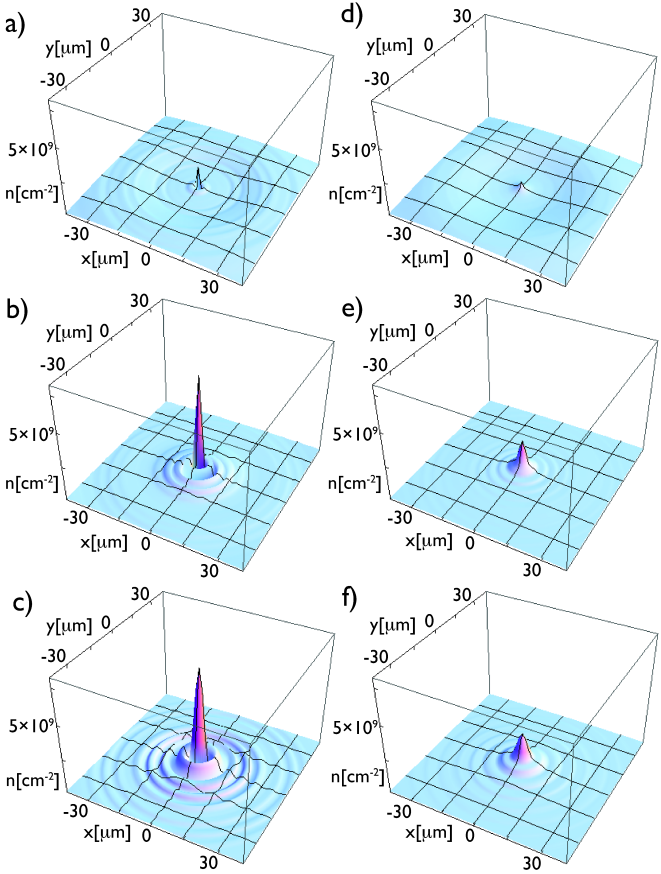}
\caption{\label{TOF} (Color online) Expectation value of the density,
  integrated over $z$ direction, after time-of-flight expansion of an
  atomic ensemble of $N_0=30000$ atoms of mass $M=23 u$, in a ring
  with radius $R = 40 \mu$m, and $\omega_z=2\pi\times 1000$Hz and
  $\omega_\rho=2\pi\times 200$Hz.  a)--c) show the TOF interference
  for $T=20$nK, at times $t=10, 20, 30$ms, d)--f) show the case of
  $T=170$nK.  }
\end{figure}

We now discuss the interference of atoms released from a ring trap.
For this purpose we use a Green's function approach which conveniently
relates the density after time-of-flight to the two-field correlation
function prior to release.  
 Assuming freely expanding atoms, the time-dependent bosonic single particle
annihilation operator is given by
\bea
\psi(\bx,t) & = & \int d\bx_1 w(\bx - \bx_1,t) \psi(\bx_1).
\eea
Here, $\psi(\bx,t)$ is the bosonic single particle annihilation operator
 at time $t$,  and $\psi(\bx_1)$ is the initial single particle
 operator.
 The Green's function of free propagation 
 $w(\bx, t)$ is 
  given by 
\bea
w(\bx, t) & = & \Big(\frac{M}{2\pi i \hbar t}\Big)^{3/2}
\exp(i M \bx^2/2\hbar t).
\eea
 The expectation of the density at time $t$ is therefore
\bea
&&n(\bx,t)  =\nonumber\\
&& \int\int d\bx_1 d\bx_2
 w^*(\bar{\bx}_1, t)w(\bar{\bx}_2, t) \langle  \psi^\dagger(\bx_1) \psi(\bx_2)\rangle\label{density}
\eea
 with $\bar{\bx}_1 = \bx - \bx_1$ and $\bar{\bx}_2 = \bx - \bx_2$.
 We are interested in the density integrated over the $z$ direction,
 and we switch to cylindrical coordinates:
\bea
n(\rho,t) & = & \int dz \, n(\rho, z, t).
\eea
 Using Eq. (\ref{density}), and integrating over $z$, we find
\bea\label{60}
n(\rho,t) &=& \Big(\frac{M}{2\pi \hbar t}\Big)^{3} \frac{\hbar t}{M}
\int  dz_1  \rho_1 d\rho_1  \rho_2 d\rho_2  d\theta_1 d\theta_2\nonumber\\
&& \exp(i M (\rho(\rho_1\cos\theta_1 - \rho_2\cos\theta_2))/\hbar t\nonumber\\
&& -i M(\rho_1^2 - \rho_2^2)/2\hbar t )
 \sqrt{n(\rho_1,z_1)n(\rho_2,z_1)}\nonumber\\
&&\exp(-R/l_\phi (|\Delta\theta| - (\Delta\theta)^2/2\pi)).
\eea
Here we assumed that the single particle correlation function
 is of the form
\bea
&&\langle\psi^\dagger(\rho_1, z_1, \theta_1)\psi(\rho_2, z_2, \theta_2)\rangle\nonumber\\
&& =\sqrt{n(\rho_1,z_1)n(\rho_2,z_1)}
\exp\Big[\frac{R}{l_\phi} \Big(\frac{\Delta\theta^2}{2\pi}-|\Delta\theta|\Big)\Big]
\eea
with $\Delta \theta = \theta_1 - \theta_2$.  Here we have ignored the
condensate depletion due to thermal activation of transverse modes.
  
We perform the integral of Eq. (\ref{60}) 
 using shifted coordinates $\tilde{\rho}_1$
and $\tilde{\rho}_2$, defined by $\rho_1 = R + \tilde{\rho}_1$ and $\rho_2 = R
+ \tilde{\rho}_2$.
 Ignoring quadratic terms in $\tilde{\rho}_{1,2}$ in the exponent we obtain
\bea\label{nrhot}
n(\rho,t) &\approx& \frac{n_{\mathrm{max}}}{4}
\int  dz_1   d\theta_1 d\theta_2 A^*(\theta_1) A(\theta_2)\nonumber\\
&& \exp(i M \rho R(\cos\theta_1 - \cos\theta_2)/\hbar t)\nonumber\\
&&\exp(-R/l_\phi (|\Delta\theta| - (\Delta\theta)^2/2\pi))
\eea
 where $A(\theta)$ is given by
\bea
A(\theta) & = & \frac{1}{(R-\rho\cos\theta)}
 \Big(R\sqrt{1-z^2_1/l_z^2}\nonumber\\
&& J_1\Big(\frac{M l_\rho \sqrt{1 - z_1^2/l_z^2}(R-\rho\cos\theta)}{\hbar t}\Big)\nonumber\\
&& + i l_\rho (1-z^2_1/l_z^2)\nonumber\\
&& J_2\Big(\frac{M l_\rho \sqrt{1 - z_1^2/l_z^2}(R-\rho\cos\theta)}{\hbar t}\Big)\Big).
\eea
We evaluate Eq. (\ref{nrhot}) numerically, for $R= 40$nm, $N_0=30000$,
$\omega_z=2\pi\times 1000$Hz, and $\omega_\rho=2\pi\times 200$Hz.  We
assume $M$ to be the sodium mass, $M=23 u$, where $u$ is the atomic
mass unit, and $a = 2.75$nm.
 We note again that in this example the interactions are taken into account
 for the initial state, whereas the time-of-flight expansion
 is ballistic. 
 The initial velocity distribution is given by
 the TF wavefunction in the transverse direction
 and by the thermal distribution
 of the phononic modes along the longitudinal direction.

 In Fig. \ref{TOF} (a) - (c) we show the case of $T = 20$nK, at
 times $t = 10, 20, 30$ms, in Fig. \ref{TOF} (d) - (f) the
 case of $T = 170$nK for the same times. 
 The latter is in the phase-fluctuating regime
 and the contrast of the expectation value of the density is visibly reduced.
  In particular, the density at the center varies with time as
\bea
n(0,t)& \approx& \frac{\pi n_{\mathrm{max}}}{2} F(l_\phi)
 \int_{-l_z}^{l_z} dz (1- z^2/l_z^2)\nonumber\\
&& J_1^2(M l_\rho \sqrt{1- z^2/l_z^2} R/\hbar t)
\eea
where
\bea
F(l_\phi) &= &  \pi \sqrt{\frac{2 l_\phi}{R}}
  \exp(-\pi R/2l_\phi)
 erf_i(\sqrt{\pi R/2 l_\phi})
\eea
is a measure of phase coherence around the ring.

For small $l_\phi/R$, $F(l_\phi)$ behaves as
 $F(l_\phi)  \approx  2 l_\phi/R$.
For large values of $l_\phi/R$, it saturates
 to $F(l_\phi) =  2\pi$.
 This behavior of the overall scale of the interference
 directly reflects the phase coherence of the BEC.
   The more coherent the system is, the higher the 
 interference contrast of the time-of-flight image.  
 
 At short times $n(0,t)$ shows a linear behavior in time,
 when time-averaged over the fast oscillatory behavior:
\bea
n(0,t)& \approx& \frac{\pi n_{\mathrm{max}}}{2} F(l_\phi)
 \frac{\hbar t l_z}{2 M l_\rho R}\label{tofshort}
\eea
At long times $n(0,t)$ falls off as $\sim 1/t^2$;
\bea
n(0,t)& \approx& \frac{\pi n_{\mathrm{max}}}{2} F(l_\phi)
 \Big(\frac{M l_\rho R}{\hbar t}\Big)^2\frac{4 l_z}{15}.\label{toflong}
\eea
 By comparing Eqs. (\ref{tofshort}) and (\ref{toflong}) 
 we can see that the time at which the center density is
 maximum scales as $t_{\mathrm{max}} \sim R/(\hbar/(l_\rho M))$.
 The radius $R$ is the distance traveled, 
 and $\hbar/(l_\rho M)$ is the scale of 
 the velocity distribution in radial direction.

 We note that the expansion of a repulsively interacting gas should be faster
 than that of a non-interacting one, because the potential energy of
 the initial state is transformed into kinetic energy.  The main
 observation, however, that a phase-coherent sample would show
 constructive interference at the center of the ring, whereas an
 incoherent sample would not, should still apply.  However, the
 increased atomic velocity would lead to smaller fringe spacings
 making optical resolution of the narrow central peak more difficult.
 For sufficiently small interactions and for high enough optical resolution,
 this method could nevertheless constitute a method
  to detect the phase-fluctuating regime.

\section{Conclusions}\label{conclude}
We have studied the phase-fluctuating regime of condensates in a
ring-shaped trap.  We first considered a simplified box geometry in
Sect. \ref{box}, and calculated the properties of the phase
correlation function.  We then turned to a realistic ring geometry in
Sect. \ref{ring}.  The main results are the estimates of the $T_\phi$
in Eq. (\ref{Tphiring}), and of $T_c$ in Eq. (\ref{Tcring}).  With these
expressions, one can determine whether the phase-fluctuating regime
can be reached, as shown in Fig. \ref{TcTphi}.
 We present a simple condition, depending only on atom number and
 trap aspect ratio, that guarantees avoidance of the phase fluctuating regime
 for all temperature below $T_c$.
 We also discussed observing the regime of phase fluctuations in time-of-flight
 images.

This work was supported by NSF under Physics Frontier Grant No.
PHY-0822671.  L.M. acknowledges support from NRC/NIST.

\def\etal{\textit{et al.}}


\begin{thebibliography}{99}
\bibitem{details}
The experimental setup will be discussed in detail elsewhere.
\bibitem{cutoff} The Bogoliubov spectrum crosses from a linear
  spectrum over to a quadratic spectrum at the inverse healing
  length. This motivates the choice of the healing length as a
  short-range cut-off.  This short range cut-off also appears in the
  long range behavior of the correlation function, as the smallest
  'unit' of the spatial variables, which is characteristic for
  low-dimenionsal systems.
\bibitem{etazeta} The parameter $\eta$ is similar to the parameter
  $\zeta$ defined for Eq.  \ref{GPEboxrs}, but includes only the
  transverse dimensions in the rescaling, whereas $\zeta$ includes all three
 dimensions, in analogy to Ref. [\onlinecite{oehberg}].
\end{thebibliography}
\end{document}